\documentclass[pra,twocolumn,showpacs,superscriptaddress]{revtex4-1}
\usepackage{amsmath,amscd,amssymb,color}
\usepackage{graphicx,amsfonts,amsthm,epsf}
\usepackage{epstopdf}
\usepackage{float}
\usepackage{hyperref}
\usepackage{enumerate}
\usepackage{array}

\usepackage{mathtools}
\usepackage{ulem}
\begin{document}
\title{Using a quantum SWAP engine to experimentally 
validate thermodynamic uncertainty relations}
\author{Krishna Shende}
\email{ph19032@iisermohali.ac.in}
\affiliation{Department of Physical Sciences, Indian
Institute of Science Education \& 
Research Mohali, Sector 81 SAS Nagar, 
Manauli PO 140306 Punjab India.}
\author{Arvind}
\email{arvind@iisermohali.ac.in}
\affiliation{Department of Physical Sciences, Indian
Institute of Science Education \& 
Research Mohali, Sector 81 SAS Nagar, 
Manauli PO 140306 Punjab India.}
\author{Kavita Dorai}
\email{kavita@iisermohali.ac.in}
\affiliation{Department of Physical Sciences, Indian
Institute of Science Education \& 
Research Mohali, Sector 81 SAS Nagar, 
Manauli PO 140306 Punjab India.}
\begin{abstract}
Thermodynamic uncertainty relations (TURs) arise from the bounds on
fluctuations of thermodynamics quantities during a non-equilibrium
process and they impose constraints on the corresponding process.  We
experimentally implement a quantum SWAP engine on a nuclear magnetic
resonance setup and demonstrate that a Gibbs thermal state can be
prepared in two different ways, either directly from a thermal
equilibrium state, or by first initializing the system in a pseudopure
state.  We show that the quantum SWAP engine can work both as a heat
engine and as a refrigerator.  Starting from a pseudopure state, we
construct the SWAP engine, and investigate the violation of two
different TURs, namely a generalized TUR and a tighter, more specific
TUR.  Our results validate that the generalized TUR is obeyed in all
the working regimes of the SWAP engine, while the tighter TUR is
violated in certain regimes.
\end{abstract} 
\maketitle 
\section{Introduction}
\label{sec1}
Substantial progress has been made in the field of non-equilibrium quantum
thermodynamics over the past two decades, which has led to an understanding of
the thermodynamic properties of systems at the microscopic
level~\cite{Goold_2016,Sai_2016}.  Recently, experimental quantum heat engines
have been implemented on different quantum
platforms~\cite{Zou_2017,Klatzow_2019}.  At the microscopic level, apart from
thermal fluctuation, quantum fluctuations also influence system dynamics and
the non-equilibrium entropy production in a quantum heat engine becomes
stochastic in nature, leading to the entropy ($\Sigma$) becoming a stochastic
variable.  These distributions follow symmetry relations, known as fluctuation
theorems (FT)~\cite{Jar1997,Crook1997,Jar_woj_2004,campisi2011,Campisi_2014},
having a general form $\frac{P(\Sigma)}{P(-\Sigma)}=e^{\Sigma}$, where
$P(\Sigma)$ and $P(-\Sigma)$ are the probability distributions for forward and
backward (time reversed) processes, respectively.  FTs have been studied using
various experimental setups such as nuclear magnetic resonance
(NMR)\cite{Serra_2014,Pal_2019,Micadei_2021,denzler_2021}, ion
traps~\cite{An_2015} and nitrogen vacancy (NV) centers\cite{Hernandez_2021}.

Another set of powerful inequalities known as thermodynamic
uncertainty relations (TURs) were formulated, which impose
restrictions on fluctuations in various thermodynamic
quantities in terms of the irreversible entropy produced
during a non-equilibrium
process~\cite{Seifert_2015,Seifert_2016}.  For a Markovian
system with $N$ discrete states, the rates of transitions
between neighboring states fulfill the local detailed
balance equation~\cite{Seifert_2016_prx}.  In such
scenarios the integrated 
thermodynamic current $Q_i$ (for instance, such
currents could arise from chemical potential differences in
biomolecular reactions or colloidal forces etc.) 
exchanged during an
out-of-equilibrium process for some definite time interval,
is bounded by the corresponding TUR (termed
TUR-1)~\cite{Seifert_2015}:
\begin{equation}
\dfrac{{\rm Var}(Q_i)}{\langle Q_i\rangle^2} 
\ge \dfrac{2}{\langle \Sigma\rangle}
\label{TUR1}
\end{equation}
where Var(Q$_i$)=$\langle Q^2_i\rangle - \langle
Q_i\rangle^2$ is the variance, and $\langle \Sigma\rangle$
is the average entropy produced during an out-of-equilibrium
process and also quantifies how far the system is driven
away from equilibrium.  TUR-1 (Eq.~\ref{TUR1}) implies a trade-off relation
between the precision i.e.  the variance in integrated
current, and the average entropy produced. Hence, to reduce
fluctuations, one must pay the inevitable price of entropy
production~\cite{Horowitz2020}.

The conditions used to derive the TUR-1 bound exclude some
key system dynamics such as systems driven by a
time-dependent protocol, microscopic heat engines, and
quantum non-equilibrium
dynamics~\cite{Ptaszynski_2018,Agarwalla_2018,Liu_2019}.
Hence, a tighter TUR was derived (termed TUR-2) assuming an
exchange fluctuation theorem (XFT)
scenario~\cite{Timpanaro_2019}: \begin{equation} \dfrac{{\rm
Var}(Q_i)}{\langle Q_i\rangle^2} \ge f(\langle \Sigma
\rangle) \label{TUR2} \end{equation} where
$f(x)=\text{cosech}^2{[g(x/2)]}$, and $g(x)$ is the inverse
function of $x \tanh{(x)}$.  In the XFT scenario, a system
can comprise $N$ subsystems, each initially prepared at
different inverse temperatures  and different chemical
potentials. These subsystems are brought in contact with
each other via a unitary operator $U$ which facilitates all
types of interactions such that the subsystems can exchange
net current, energies and particles with each other.  TURs
have been recently studied experimentally in the heat
exchange case~\cite{TSM_prr}, where two qubits initially
prepared at different spin temperature interacted via an
external driving unitary, such that heat is exchanged
between the qubits. 
A series expansion of $f(\langle \Sigma \rangle )$ around $\langle \Sigma
\rangle=0$ gives $f(\langle \Sigma \rangle )\approx 2/{\langle \Sigma \rangle}
-2/3$\cite{Timpanaro_2019}. Consequently, TUR-2 (Eq.~\ref{TUR2}) simplifies to
TUR-1 (Eq.~\ref{TUR1}) when entropy values are minimal($\langle \Sigma \rangle
\rightarrow 0)$).

Quantum engines can be used to examine various thermodynamic properties at the
quantum level  and three types of quantum engines have been proposed in the
literature~\cite{Quan_2007,Raam_2015}: (i) a 4-stroke heat engine also known as
Otto engine, (ii) a 2-stroke heat engine also called the SWAP engine, and (iii)
a continuous heat engine.  Different types of quantum engines have been
experimentally implemented on NMR quantum
processors~\cite{Peterson_2019,Assis_2019,Lisboa_2022} NV
centers~\cite{NV_engine_2019}, in a trapped-ion setup~\cite{Trap1,Trap2} and
using superconducting qubits~\cite{PRX_IBM}.  The smallest possible
refrigerator was recently built using three NMR qubits which operates without
relying on net external work~\cite{fridge-daweilu-prl-2024}.  A SWAP quantum
engine was implemented on a cloud quantum processor which has a boosted
efficiency above the standard Carnot limit~\cite{herrera-prr-2023}.

In this work, we experimentally demonstrate the implementation of a quantum
SWAP engine~\cite{Campisi_2015} on a two-qubit NMR quantum processor.  We show
that the quantum SWAP engine can be constructed in two different ways, by
starting from a thermal equilibrium state, or by first preparing the system in
a pseudopure state.  We observe the working of the SWAP engine as a heat engine
and as a refrigerator, depending on the spin temperature and the energy gap of
the qubits.  The spin temperature of one qubit is held fixed at two different
values and the spin temperature of the other qubit is changed in order to probe
the dynamics of the engine.  We use the quantum SWAP engine to explore the
validity of the TUR-1 and TUR-2 bounds  in different operation modes  of the
quantum SWAP engine.

This paper is organized as follows: The theoretical
framework of the quantum SWAP engine is explained in
Section~\ref{sec2}, with a detailed calculation of its
working principle and required quantities.
Section~\ref{sec3} describes the NMR experimental setup and
contains details about the implementation of the quantum
SWAP engine on a two-qubit NMR platform. Section~\ref{sec4}
contains a discussion of the experimental
study of the TURs using a quantum SWAP engine.
Section~\ref{sec5} offers some concluding remarks.
\section{Theoretical Background}
\label{sec2}
We briefly describe the operational aspects of a quantum SWAP
engine and its implementation on an NMR quantum processor and also
provide definitions of the quantities required for computing 
the TURs. 
\subsection{NMR SWAP Engine}
\label{sec2b}
\begin{figure}[h]
\includegraphics[scale=1]{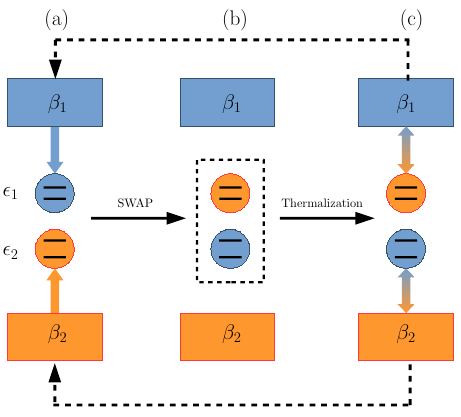}
\caption{Schematic diagram showing operation of 
the quantum SWAP engine in the heat engine mode.
(a) The two qubits are thermalized to their respective baths, with
$\beta_1\ge\beta_2$ and $\epsilon_2>\epsilon_1$. (b) Both the qubits are
decoupled from their baths and allowed to interact 
with each other via a SWAP operation. (c)
The qubits are once again thermalized with their
respective baths, to begin a new thermodynamic cycle.}
\label{swap_engine}
\end{figure}

Consider two NMR qubits (denoted by qubit 1 and qubit 2)
with an energy level spacing of $\epsilon_1$ and $
\epsilon_2$, respectively.  Two heat baths are initialized
to different spin temperatures, such that
$\beta_2\le \beta_1$, i.e.  the first bath is colder than
the second.  A schematic diagram of this two-qubit SWAP
engine is shown in Figure~\ref{swap_engine}, with different
energy gaps and initial temperatures.  The working of this
engine is carried out in two steps, termed as the first and
second stroke, respectively (shown in
Figures~\ref{swap_engine}(a) and (b)).

\noindent{\bf First stroke of the SWAP engine}: During the
execution of the first stroke of the engine, the qubits with
Zeeman Hamiltonians $H_1=-h \epsilon_1 I_z$ and $H_2=-h
\epsilon_2 I_z$ are initialized to  different temperatures
via equilibration with their respective heat baths.  The
composite system is in a diagonal state given by
$\rho_i=exp[-\beta_iH_i]/Z_i$, which is a Gibbs state given
by\cite{Campisi_2014}:

\begin{equation}
\rho^{(a)}=\rho_1^{(a)} \otimes
\rho_2^{(a)}=e^{\frac{-\beta_1H_1}{Z_1}}\otimes
e^{\frac{-\beta_2H_2}{Z_2}} 
\label{initial_gibbs_state}
\end{equation}
where the superscript $a$ refers to the first stroke of the
engine, and $\beta_i=1/(k_B T_i)$ and $Z_i={\rm
Tr}[\exp{(-\beta_i H_i)}]$ are the inverse spin temperature
and the partition function of the $i^{th}$ qubit,
respectively.

The initial density operator of each qubit is given by:
\begin{equation}
\begin{split}
\rho_1^{(a)} =
\frac{1}{Z_1}\{|0\rangle_1\langle0|_1+e^{-\beta_1\epsilon_1}|1
\rangle_1\langle1|_1\}\\
  \rho_i^{(a)} =
  \frac{1}{Z_2}\{|0\rangle_2\langle0|_2+e^{-\beta_2\epsilon_2}|1
  \rangle_2\langle1|_2\}
\end{split}
\end{equation} 
where $\beta_{1(2)}=1/k_BT_{1(2)}$ and for simplicity the eigen energy
of the ground state has been chosen as a reference
point~\cite{Quan_2007}.

\noindent{\bf Second stroke of the SWAP engine}: 
During the second stroke of the SWAP engine, both qubits are disconnected from
their heat baths, and the qubits are allowed to interact via a quantum SWAP
unitary $U$ gate given by: 
\begin{equation}	
U=\begin{bmatrix}
1 & 0 & 0 & 0 \\
0 & 0 & 1 & 0 \\
0 & 1 & 0 & 0 \\
0 & 0 & 0 & 1 \\
\end{bmatrix}  
\label{3q_matrix}
\end{equation} 
which exchanges the states of both the qubits~\cite{nielsen_chuang_2010}.  The
density operator for each qubit after the SWAP gate operation is given by:
\begin{equation}
\begin{split}
\rho_1^{(b)} =
\frac{1}{Z_2}\{|0\rangle_1\langle0|_1+e^{-\beta_2\epsilon_2}
|1\rangle_1\langle1|_1\}\\
\rho_2^{(b)} =
\frac{1}{Z_1}\{|0\rangle_2\langle0|_2+e^{-\beta_1\epsilon_1}
|1\rangle_2\langle1|_2\}
\end{split}
\end{equation}
where the superscript $b$ refers to the second stroke of the engine.
After this, the qubits are once again allowed to equilibrate with their
respective heat baths, in order to begin a new thermodynamic 
cycle (Figure~\ref{swap_engine}(c)).  The change
in the energy of each qubit is given by:
\begin{equation}
<\Delta E_i>={\rm Tr}[H_i(\rho^{(b)}-\rho^{(a)})]
\end{equation}
where $H_i$ is the Hamiltonian of $i^{th}$ qubit. The average work done is given by: 
\begin{equation}
\langle W \rangle= \langle \Delta E_{1} \rangle + \langle \Delta E_{2} \rangle
\end{equation}

The first stroke of the SWAP engine wherein both qubits are thermalized, is
equivalent to two quantum isochoric processes, while the SWAP operation which
is implemented during the second stroke of the SWAP engine can be regarded as
two quantum adiabatic steps, when the work is done. During the
thermalization process, the energy change in each  
individual qubit due to the second
stroke of the engine is dumped in its respective heat 
bath. This energy change can be
related to the heat flow of the individual qubits 
as $Q_1 =-\langle \Delta E_1 \rangle$ and
$Q_2 =-\langle \Delta E_2 \rangle$.

The heat absorbed by qubit 1 is given by\cite{Quan_2007,Campisi_2015}:
\begin{equation}
\begin{aligned}
Q_{1}&= -<\Delta E_1>={\rm Tr}[H_1(\rho_1^{(b)} - \rho_1^{(a)})].
\label{Qout}\\
&=-\epsilon_1\left[\frac{e^{-\beta_2\epsilon_2}}{Z_2}-
\frac{e^{-\beta_1\epsilon_1}}{Z_1}\right]
\end{aligned}
\end{equation}

The heat released by qubit 2 is given by~\cite{Quan_2007,Campisi_2015}:
\begin{equation}
\begin{aligned}
Q_{2}&=-<\Delta E_2>={\rm Tr}[H_2(\rho_2^{(b)} - \rho_2^{(a)})],
\label{heat}\\
&=-\epsilon_2\left[\frac{e^{-\beta_1\epsilon_1}}{Z_1}-
\frac{e^{-\beta_2\epsilon_2}}{Z_2}\right]\\
&=\epsilon_2\left[\frac{e^{-\beta_2\epsilon_2}}{Z_2}-
\frac{e^{-\beta_1\epsilon_1}}{Z_1}\right] 
\end{aligned}	
\end{equation}

The extracted work is given by\cite{Quan_2007,Campisi_2015}:
\begin{equation}
\begin{aligned}
W&=<\Delta E_{1}>+<\Delta E_{2}>=-(Q_1+Q_2)\\
&=(\epsilon_2-\epsilon_1)\left[\frac{e^{-\beta_2\epsilon_2}}{Z_2}-
\frac{e^{-\beta_1\epsilon_1}}{Z_1}\right]\label{work}
\end{aligned}
\end{equation}
The difference in the average energy change of each qubit gives the work, and
also  gives the direction of heat flow.  The system can work 
either as an engine or as a
refrigerator depending on whether the heat flows from the hot to the cold qubit
or vice versa.  The quantity  Q$_i$ is positive if heat flows out from a qubit
and negative if heat is gained by the qubit.  It can be inferred from the above
equations that, if both the qubits have the same energy gap, the work done is
always zero. Hence, for any quantum system to work as a SWAP engine, both the
qubits must have different energy gaps.

The quantum SWAP heat engine efficiency is computed to be\cite{Campisi_2015,Timpanaro_2019}:
\begin{equation}
\eta=-\frac{W}{Q_{2}}=1-\frac{\epsilon_1}{\epsilon_2}
\label{effi}
\end{equation}

When heat flows from a cold to a hot reservoir, the system works as a
refrigerator(W $>$ 0) and work is done on the system.
The system works as a heat engine when work (W$<$0) is extracted
from it.  The conditions for the working of a quantum SWAP engine are delineated
below~\cite{Campisi_2015,Timpanaro_2019}:
\begin{itemize}
\item Heat engine: 
$1 > \frac{\epsilon_1}{\epsilon_2}>\frac{\beta_2}{\beta_1}$
\item Refrigerator: 
$\frac{\beta_2}{\beta_1} > \frac{\epsilon_1}{\epsilon_2} > 0$
\end{itemize}

\subsection{Average Entropy and variance}
The central quantities required to explore TURs (Eq.~\ref{TUR1} and
Eq.~\ref{TUR2}) in different working regimes of the quantum SWAP engine are the
variances of heat and work, 
and the average entropy produced when the system goes
to a non-equilibrium state.  A TUR places a restriction on the uncertainty
(variance) in the integrated current in terms of entropy produced. Hence, if one
wants to decrease this uncertainty, one has to pay the inevitable price of more
entropy production.

The average entropy produced by a SWAP engine is given 
by~\cite{Timpanaro_2019}: 
\begin{equation}
\langle \Sigma \rangle= (\beta_1 - \beta_2)\langle Q 
\rangle -\beta_1\langle W \rangle
\label{avg_entropy}
\end{equation} 
where $\langle Q \rangle$ and $\langle W \rangle$ are the average released heat
and average extracted work as given in Eq. \ref{heat} and Eq. \ref{work},
respectively. We denote the heat released from qubit 2 (Q$_2$)
as Q. 
The variance of any quantity denotes
the deviation from its mean value. 
The variances
of heat and work and all their cumulants 
can be computed from the cumulant 
generating function to be~\cite{Timpanaro_2019}:
\begin{widetext}
	\begin{equation}
	Var(Q)/2=	\dfrac{2\epsilon_1^2 e^{\beta_2 \epsilon_2}}{(1+e^{\beta_2 \epsilon_2})(1+e^{\beta_1 \epsilon_1})}+\dfrac{\epsilon_1 Q_2}{1+e^{\beta_1 \epsilon_1}}\\
	+\dfrac{\epsilon_1 Q_2 e^{\beta_2 \epsilon_2} }{1+e^{\beta_2 \epsilon_2}}
	\label{var_Q}
	\end{equation}

	\begin{equation}
	\begin{split}
	Var(W)/2=&\dfrac{2(\epsilon_1 e^{\beta_1 \epsilon_1}+\epsilon_2)(\epsilon_2 e^{\beta_2\epsilon_2}+\epsilon_1))}{(1+e^{\beta_2 \epsilon_2})(1+e^{\beta_1 \epsilon_1})}-
	\dfrac{(\epsilon_1 e^{\beta_1 \epsilon_1}+\epsilon_2)W}{1+e^{\beta_1 \epsilon_1}}-\dfrac{(\epsilon_2 e^{\beta_2 \epsilon_2}+\epsilon_1)W}{1+e^{\beta_2 \epsilon_2}}\\
	&+\dfrac{\epsilon_1^2 e^{\beta_1 \epsilon_1} +\epsilon_2^2}{1+e^{\beta_1 \epsilon_1}}+\dfrac{\epsilon_2^2 exp(\beta_2 \epsilon_2) +\epsilon_1^2}{1+e^{\beta_2 \epsilon_2}}-(\epsilon_1 +\epsilon_2)\dfrac{(\epsilon_1e^{\beta_1 \epsilon_1}+\epsilon_2)}{1+e^{\beta_1 \epsilon_1}}-(\epsilon_1 +\epsilon_2)\dfrac{(\epsilon_2e^{\beta_2 \epsilon_2}+\epsilon_1)}{1+e^{\beta_2 \epsilon_2}}
	\end{split}
	\label{var_w}
	\end{equation}
\end{widetext}
where $\beta_1$ and $\beta_2$ are the 
inverse spin temperatures of qubit 1 and qubit
2 and $\epsilon_1$ and $\epsilon_2$ are the energy gaps of 
qubit 1 and qubit 2, respectively. 

Recently,  a novel identity connecting the
inverse signal-to-noise ratio between heat and work has been identified,
specifically in the context of the SWAP engine~\cite{Sacchi_2021}:
\begin{equation}
	\dfrac{{\rm Var}(Q)}{\langle Q \rangle^2}=	
	\dfrac{{\rm Var}(W)}{\langle W \rangle^2}
\label{equal}
\end{equation}
which implies that the inverse signal-to-noise ratio of heat and work are
equal. The difference between the average value of work and heat is compensated
for by having more uncertainty in either work or heat, such that the inverse
signal-to-noise ratio of heat and work remains the same. All the above
relations (Eqn.~\ref{Qout}-\ref{equal}) 
capture the interdependence among heat,
work and entropy production.
\section{Experimental implementation}
\label{sec3}
\begin{figure}[h]
\includegraphics[scale=1]{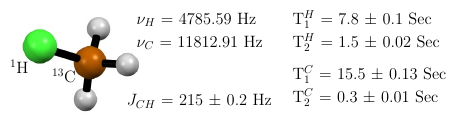}
\caption{Molecular structure of $^{13}$C-labeled CHCl$_3$ with the two qubits
being realized by the $^1$H and $^{13}$C spins.  The offset rotation frequencies
($\nu$), the scalar coupling strength ($J$) and the  T$_1$ and T$_2$ relaxation
times for each qubit are shown alongside.}
\label{molecule-figure}
\end{figure}

A $^{13}$C-labeled chloroform molecule dissolved in Acetone-D6 was used to
realize a two-qubit system, where the $^1$H and $^{13}$C spins are encoded as
qubit~1 and qubit~2, respectively (the molecular structure and experimental
parameters are depicted in Figure~\ref{molecule-figure}).  Experiments were
performed on a Bruker Avance III 600 MHz NMR spectrometer equipped with a 5~mm
TXI probe.  The NMR Hamiltonian for a two-qubit qubit system is given
by~\cite{Chunag_review,OLIVEIRA}:
\begin{equation}
\mathcal{H}=-\sum\limits_{i=1}^2 
{ \frac{h\nu_i}{2} } \sigma_z^i + \sum\limits_{i<j=1}^2 
\frac{hJ_{ij}}{4} \sigma_z^i \sigma_z^j
\label{nmr-hamil}
\end{equation}
where $\nu_i$ the offset frequency, $\sigma_z^i$  is the $z$ component of the
spin angular momentum and J$_{ij}$ is the scalar coupling between spins $i$ and
$j$, respectively. The Hamiltonians $H_1$ and $H_2$ in
Eq.~\ref{initial_gibbs_state} 
can be realized from the
Hamiltonian in Eq.~\ref{nmr-hamil} as $H_1=\frac{-h\nu_1}{2}  \sigma_z^1$ and
$H_2=\frac{-h\nu_2}{2}  \sigma_z^2$, where $\nu_1$=4785.59 Hz and
$\nu_2$=11812.91 Hz, respectively.

\begin{figure}[h]
\includegraphics[scale=1]{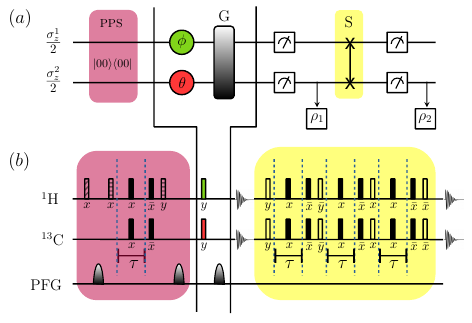}
\caption{(a) Quantum circuit to implement a two-qubit quantum SWAP engine.  The
first box represents PPS state preparation.  The green and red circles represent
transverse rf pulse along the $y$-axis and the values of $\theta$ and $\phi$
control the spin population (which is directly proportional to its spin
temperature).  (b) NMR pulse sequence to realize the quantum SWAP heat engine.
Bars with diagonal lines, bars with horizontal lines, unfilled bars and bars
having crosshatched lines represent rf pulses of flip angle $(\frac{\pi}{3})$,
$(\frac{\pi}{4})$, $(\pi)$ and $(\frac{\pi}{2})$, respectively.  The phases of
the rf pulses are written below their respective bars.  Grey and black bars
correspond to pulses of angles  $(\phi)$ and $(\theta)$, respectively.  The line
denoted by `PFG' depicts the times at which a pulsed field gradient magnetic
pulse is employed to destroy coherence.  The time interval $\tau$ is set equal
to $\frac{1}{2{\rm J}_{{\rm CH}}}$.} 
\label{ckt_diag}
\end{figure}

The quantum circuit to implement the SWAP engine is shown in
Figure~\ref{ckt_diag}. Single-qubit rotation gates are implemented via
high-power rf pulses of short duration.  The system is allowed to freely evolve
under the Hamiltonian during time periods when no pulses are applied; evolution
periods are interspersed with $\pi$ pulses to refocus chemical shifts and retain
only the desired scalar coupling interactions.  For $^{1}$H and $^{13}$C spins,
the duration of $\dfrac{\pi}{2}$ pulses are 7.2 $\mu$s and 12.80 $\mu$s at power
levels of 19.9 W and 237.3 W,  respectively.  The spin-lattice and spin-spin
relaxation times of both qubits are much longer than the gate implementation
times and hence do not affect the overall experimental performance.

The implementation of the quantum SWAP engine can be divided into two parts:
(i) the initial thermalization into the Gibbs state and (ii) the SWAP operation
during which the work is done.  The effective spin temperature($\beta_i$) of
the initial Gibbs state is related to the ground ($p_0$) and excited ($p_1$)
populations as~\cite{Quan_2007}:
\begin{equation} 
\beta_i=\dfrac{1}{h\nu} \ln \left(\frac{p_0}{p_1}\right)
\end{equation} 
where $h$ is Planck's constant and $\nu$ is the offset frequency. 

We have implemented the first stroke of the engine using two different schemes,
which we call the (i) Direct SWAP engine and  the (ii) PPS SWAP engine,
respectively.  The circuit and pulse sequence to realize the direct SWAP engine
are shown in Figure~\ref{ckt_diag}(a) and (b), respectively.  The circuit and
pulse sequence inside the pink shaded boxes are not implemented and only the
circuit and pulse sequence within the two black vertical lines are used.  The
magnetization strength is varied by applying strong rf pulses along the
$y$-axis, of flip angles ranging between 0$^{\circ}$ to 90$^{\circ}$, on both
the qubits.  The angle of rotation of these pulses controls the ratio of the
population between the ground and excited states.  This leads to a
redistribution of populations in the energy states of the spins and the
effective spin temperature of the nuclei is changed accordingly.  Any residual
coherence in the system is killed by applying pulsed field gradient pulses
along the $z$-axis.  Quantum state tomography is then performed using the
constrained convex optimization approach~\cite{Gaikwad2021}, to reconstruct the
density operator $\rho_1$, corresponding to the initial Gibbs state.  

In the scheme to implement the PPS SWAP engine,
the initial thermalization to the Gibbs state at different spin
temperatures is achieved by first preparing the system in a 
pseudopure state (PPS)
using the spatial averaging technique and pulsed field gradients:
\begin{equation}
\rho_{{\rm PPS}}=\frac{1}{4}(1-\eta)I+\eta|00\rangle\langle00|.
\end{equation}
where $\eta$ is a factor proportional to the spin polarization which
depends on the external magnetic field and the gyromagnetic
ratio.
The initial thermalization is achieved by
implementing the circuit before the first vertical black
line in Figure~\ref{ckt_diag}(b).  
The rest of the PPS SWAP engine implementation proceeds as described
above for the direct SWAP engine.

The second stroke of the engine consists of the quantum SWAP gate and is
implemented experimentally using the part of the NMR pulse sequence after the
second vertical black line in Figure~\ref{ckt_diag}(b).  Quantum state
tomography is once again performed to reconstruct the final density operator
$\rho_2$, after the second stroke of the engine.  This completes one full
thermodynamic cycle of the heat engine.  A new thermodynamic cycle is begun by
letting each spin relax to its thermal equilibrium state by waiting 6.5$\times
T_1$ of the spins.

The density operator $\rho_1$ and $\rho_2$ are used to
evaluate average heat, work and $\Sigma$ as given in
Eq.\ref{heat}, Eq.\ref{work} and Eq.\ref{avg_entropy}.  The
closeness between the theoretically predicted density
operator ($\rho_t$) and the experimentally reconstructed density
operator ($\rho_e$) is characterized by the
fidelity~\cite{Zhang_f}:
\begin{equation}
F=\frac{|Tr(\rho_e
\rho_t^{\dagger})|}{\sqrt{Tr(\rho_e\rho_e^{\dagger})Tr(\rho_t\rho_t^{\dagger})}}
\end{equation}
All the experimental states were reconstructed with 
an excellent fidelity of  $\approx 99 \%$.
\section{Results and discussion}
\label{sec4}
\subsection{Direct SWAP Engine Dynamics}
\begin{figure}
\includegraphics[scale=1]{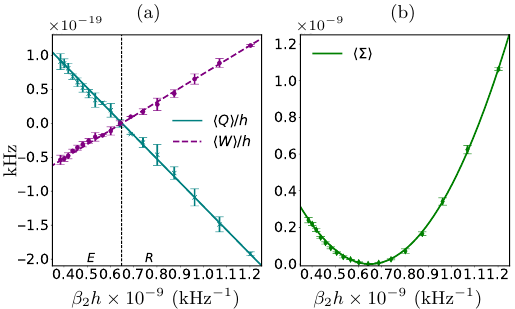}
\caption{Plots depicting dynamics of the direct SWAP engine. (a) Average heat
and work values for different spin temperatures of the second qubit,
keeping the value of the inverse spin temperature of the first qubit
fixed at $\beta_1h=1.61\times 10^{-10}$ kHz$^{-1}$.  Theoretically
calculated average values of heat and work are shown as solid teal and dashed
purple lines, respectively.  The experimentally obtained values (with
error bars) are shown as teal crosses and purple circles, respectively.
The SWAP engine functions as a heat engine in the region to the left of
the dashed vertical line and as a refrigerator in the region to the
right of the dashed vertical line.  (b) Average value of entropy
($\langle\Sigma\rangle$).  The theoretical prediction is depicted by a
solid green line and the experimental values (with error bars) are
shown as green diamonds. 
} 
\label{thermal_plots} 
\end{figure}

The direct SWAP engine was experimentally realized using the pulse sequence
given in Figure~\ref{ckt_diag}(b) (skipping elements in the pink colored box).
The spin temperature of the first qubit was kept fixed at $\beta_1h=1.61\times
10^{-10}$ kHz$^{-1}$ (300 K), while the spin temperature of second qubit was
varied from $\beta_2h=1.2\times 10^{-10}$ kHz$^{-1}$(400 K) to
$\beta_2h=1.25\times 10^{-9}$ kHz$^{-1}$(1200 K).  The plots for entropy
production, heat and extracted work are shown in Figure~\ref{thermal_plots}.
It is evident from the plots that for certain values of spin temperature, the
direct SWAP engine works as a heat engine and as a refrigerator for other spin
temperatures.  These working regimes  are separated by the black dashed vertical
line as shown in Figure~\ref{thermal_plots}(a).  Theoretically computed average
heat and work are shown as solid teal and dashed purple lines in
Figure~\ref{thermal_plots}(a), while the experimentally obtained values (along
with error bars) are depicted as teal crosses and purple circles, respectively.
There is a very good agreement between the theoretical and experimental values.
The average entropy after each thermodynamic cycle for a given value of spin
temperature of both the qubits is shown in Figure~\ref{thermal_plots}(b).  The
solid green curve is the theoretically predicted value while the green diamonds
are the experimentally obtained values of average entropy.  The average entropy
becomes zero when the condition $
\frac{\epsilon_1}{\epsilon_2}=\frac{\beta_2}{\beta_1}$ is satisfied. For this
condition, the average heat and work becomes zero, which leads to zero entropy
production. If we move away from this condition, the average entropy production
is non-zero and it increases depending upon how far the system is from
equilibrium. We observed that $\langle \Sigma \rangle \ge 0$, which follows
from the second law of thermodynamics.  As observed in
Figure~\ref{thermal_plots}(b), the average entropy produced by the direct SWAP
engine is of the order of $10^{-9}$.  For such low entropy values, the inverse
signal-to-noise ratio and the RHS of TUR-1(Eq.~\ref{TUR1}) 
and TUR-2(Eq.~\ref{TUR2}) become equal.

\subsection{PPS SWAP Engine Dynamics}
The direct SWAP engine implementation method 
offers the advantage that the nuclei are
already in the Gibbs thermal state, which can be used as an initial state.
Furthermore, it can be calculated that using the direct SWAP engine method,
$\approx 6.023 \times 10^{23}$ spins participate in the engine dynamics whereas
using the PPS SWAP engine, a much lower number $\approx 7.287 \times 10^{18}$
spins participate in the engine dynamics.  While the engine dynamics in
different working regimes  can be easily captured by using the direct SWAP
engine, the entropy produced by this engine is very low. Our numerical
calculations showed that violation of TUR-1 occurs for spin temperature values
lower than 0.266 $\mu K$ for the second qubit.  Hence, despite the easier
implementation protocol of the direct SWAP engine, we decided to use the PPS
SWAP engine to study the violation of TUR-1 and TUR-2.

\begin{figure}
\includegraphics[scale=1]{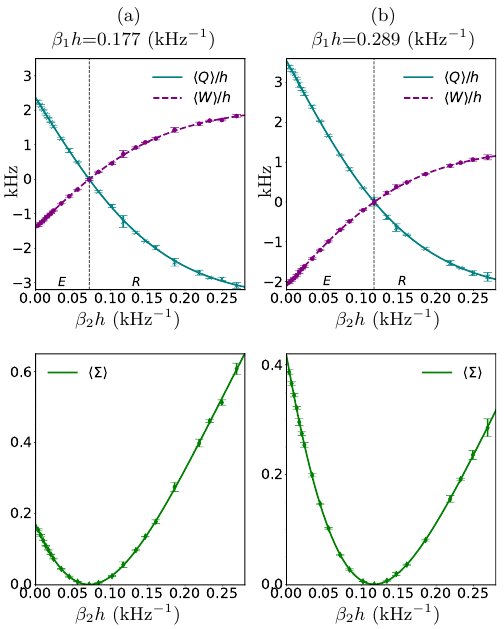}
\caption{Plots depicting dynamics of the PPS SWAP engine. Top plots in panels
(a) and (b) depict average heat and work values for different spin
temperatures of the second qubit, keeping the value of the inverse spin
temperature of the first qubit fixed at $\beta_1h=0.177$ kHz$^{-1}$ and
$\beta_1h=0.289$ kHz$^{-1}$, respectively.  Theoretically calculated
average values of heat and work are shown as solid teal and dashed
purple lines, respectively.  The experimentally obtained values (with
error bars) are shown as teal crosses and purple circles, respectively.
The SWAP engine functions as a heat engine in the region to the left of
the dashed vertical line and as a refrigerator in the region to the
right of the dashed vertical line.  Bottom plots in panels (a) and (b)
depict the average value of entropy ($\langle\Sigma\rangle$) at the two
fixed inverse spin temperature of the first qubit.  The theoretical
prediction is depicted by a solid green line and the experimental
values (with error bars) are shown as green diamonds.
}
\label{engine-plot}
\end{figure}

To study the dynamics of the PPS SWAP engine and explore the validity of the
TURs, we keep the inverse spin temperature $\beta_1 h$ of the first qubit fixed
at two different values namely, at 0.177(kHz$^{-1}$)=0.271 $\mu K$ and at
0.289(kHz$^{-1}$)=0.165 $\mu K$.  The inverse spin temperature $\beta_2h$ of
second qubit is varied, which allows us to investigate the engine dynamics at
different spin temperatures.  We plotted the average irreversible entropy
$\langle\Sigma\rangle$, average work $\langle W\rangle$ and heat $\langle
Q\rangle$ as a function of $\beta_2h$ in Figure~\ref{engine-plot}.  From the
top plots of Figure~\ref{engine-plot}, it can be inferred that
$\langle\Sigma\rangle \ge$ 0, which can be regarded as recasting of the second
law of thermodynamics for the stochastic regime at the microscopic level. This
is a confirmation of the second law for a non-equilibrium thermodynamic
process. When the energy gap ratio of both qubits  is the same as their spin
temperature ratio ( $\frac{\epsilon_1}{\epsilon_2}= \frac{\beta_2}{\beta_1}$),
the average heat and work are zero, and hence the average entropy change is
also zero.  At this point, both the work and heat curves cross each other.  We
obtain a very good agreement between our experimental values and the
theoretical predictions as shown in Figure~\ref{engine-plot}. 

\begin{figure}[h!]
\includegraphics[scale=1]{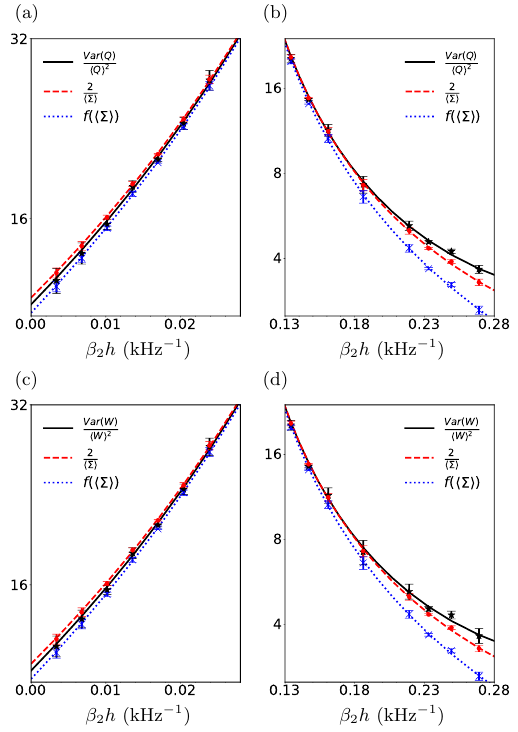} 
\caption{Comparison between TUR bounds. The  $\beta_1h$ value of the first
qubit is kept fixed at 0.177 (kHz$^{-1}$) and spin temperatures of the
second qubit are plotted on the $x$-axis.  Inverse noise-to-signal
ratio, TUR-1 and TUR-2 are depicted with a solid black, a dashed red
and a dotted blue line, respectively and their experimental values
(with error bars) are shown with black stars, red circles and blue
crosses, respectively.  Plots (a) and (c) depict the heat engine regime
while plots (b) and (d) depict the refrigerator regime for heat and
work of inverse noise-to-signal ratio.  A violation of TUR-1 is
observed in the heat engine regime, while TUR-2 is always valid in
every parameter space.  Inverse noise-to-signal ratio is the same for
heat and work, which can be observed by comparing plot (a) with (c) and
plot (b) with (d).
} 
\label{TUR_plt1}
\end{figure}

\begin{figure}[h!] 
\includegraphics[scale=1]{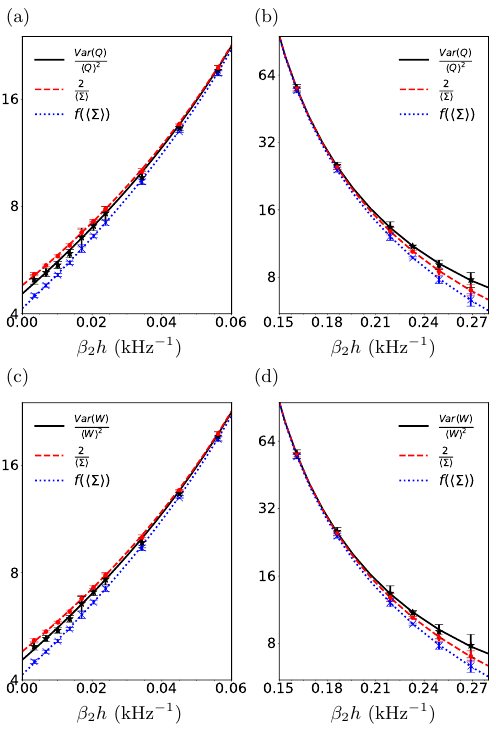}
\caption{Same as Figure~\ref{TUR_plt1}, but with the spin temperature of
qubit 1 kept fixed at $\beta_1h$=0.289 (kHz$^{-1}$).}
\label{TUR_plt2} \end{figure}

Work is extracted ($\langle W\rangle<0$) from the quantum SWAP engine and it
functions as a heat engine when $\beta_2<(\epsilon_1/\epsilon_2)\beta_1$. This
can be observed from the bottom plots of Figure~\ref{engine-plot}. The
condition for the engine to work as a refrigerator is
$\beta_2>(\epsilon_1/\epsilon_2)\beta_1$, where work is consumed ($\langle
W\rangle>0$) from an external agent to make heat flow from a cold to a hot
bath.  The parameter space for the SWAP engine working as a heat engine and as
a refrigerator is divided by a black dashed line shown in the bottom plots of
Figure~\ref{engine-plot}. 

The ratios of the variances of heat and work to the square of their average
values (LHS of TUR-1 and TUR-2) are plotted in Figures~\ref{TUR_plt1} and
\ref{TUR_plt2} as solid black lines.  This inverse signal-to-noise ratio is
compared with the TUR-1 and TUR-2 bounds (depicted as dotted red and dashed
blue curves, respectively).  The experimental data points (with error bars) are
shown in the same colors as their respective theoretical predictions.  For
small entropy production limits ($\langle \Sigma\rangle \approx 0$) the TUR-2
bound reduces to the TUR-1 bound, which can be seen as all three lines are
merging into one single line as we move towards the $\beta_2h$ values which
have nearly zero entropy.  As evident from Figures~\ref{TUR_plt1} and
\ref{TUR_plt2}, TUR-2 is always satisfied as it is derived from the XFT
scenario (which includes the quantum SWAP engine).  A violation of the TUR-1
bound is observed in the heat engine regime.  The violation of TUR-1 is
directly proportional to the entropy production in the SWAP engine.  When the
average entropy production is increased in the heat engine regime, the
violation becomes more significant (the dotted red curve is above the solid
black curve).  In the refrigerator regime, it is observed that both the bounds
weaken as the entropy production is increased.  This could open up the
possibility of a new TUR which is tighter than TUR-2 for XFT cases. 
\section{Conclusions}
\label{sec5}
We used an NMR quantum processor to implement a quantum SWAP engine using two
different methods, namely, a direct method and a PPS method.  The direct method
is easier to implement experimentally and exploits the fact that the spins are
already in the Gibbs thermal state, which itself can then be used as the
initial thermal state.  The quantum SWAP engine was characterized in its
working regimes as an engine and a refrigerator, depending on its initial
equilibrium state.  

Further, we explored the validity of the TURs using a quantum SWAP engine
implemented via the PPS SWAP engine method.  Our results indicate that the
irreversible entropy production is always nonzero, which follows the second law
of thermodynamics at the stochastic level.  We obtained a very good match
between the theoretical calculations and the experimental results, which
validates using NMR quantum processors as heat engines at the microscopic
level.  Two different bounds on inverse noise-to-signal ratio, which were
derived using different physical scenarios, were explored.  We experimentally
studied these bounds and verified that TUR-2 is always valid in every working
parameter space of the quantum SWAP engine, while TUR-1 is violated in certain
working regimes of the quantum SWAP engine.  Our study is a step forward in the
emerging field of experimental explorations of non-equilibrium quantum
thermodynamics.
\begin{acknowledgments}
All experiments were performed on a Bruker Avance-III 600
MHz NMR spectrometer
at the NMR Research Facility at IISER Mohali.  
K.~S. acknowledges financial support from the
Prime Minister’s Research Fellowship (PMRF) scheme of the
Government of India.
\end{acknowledgments}

%
\end{document}